\documentclass[aps,prl,twocolumn,groupedaddress]{revtex4-1}
\usepackage{graphicx}
\usepackage{epsfig}
\usepackage{amsmath}
\usepackage{color}
\begin{document}

\title{Violation of Geometrical Scaling in pp Collisions at NA61/SHINE}
\author{Michal Praszalowicz}
\affiliation{M. Smoluchowski Institute of Physics, Jagellonian University, Reymonta 4,
30-059 Krakow, Poland}

\date{\today}

\begin{abstract}
We analyze geometrical scaling (GS) of negative pion multiplicity $p_{\rm T}$
distributions at NA61/SHINE energies. We show that even though NA61/SHINE 
energies are low, one may expect to find GS in the particle spectra.
We argue that qualitative behavior of ratios of
multiplicities at different energies is in agreement with a
simple picture of GS which is violated for $p_{\rm T}$ smaller than some
nonperturbative scale $\Lambda$ and when larger Bjorken $x$ of  one of the
scattering patrons crosses $x_{\rm max}$ above which gluonic cloud
becomes dilute and quark degrees of freedom become important.
\end{abstract}


\maketitle

Scaling properties of physical observables have been for a long time one of
the most instructive ways to get information about the underlying physics. In
some cases this information could be fully quantified; here the most prominent
example is Bjorken scaling of Deep Inelastic Scattering (DIS) structure 
functions \cite{Bjorken:1968dy}. In some other cases, like the Koba -Nielsen-Olesen 
scaling 
\cite{Koba:1972ng} for example, the underlying physics is
still poorly understood. Some theoretically conjectured scaling laws, like
Feynman scaling in high energy hadronic collisions \cite{Feynman:1969ej}, have not been
experimentally confirmed. In any case both the emergence of a scaling law, and
-- sometimes more importantly -- its violation, test our understanding of
microscopic phenomena that are of importance in given experimental conditions.
Again Bjorken scaling is here a benchmark, since both its emergence and
violation are fully understood in terms of perturbative quantum 
chromodynamics (QCD) \cite{DGLAP}.

In this paper we will focus on another scaling law, called geometrical scaling
(GS), which has been introduced in the context of DIS \cite{Stasto:2000er}. It
has been shown that GS scaling is present also in high energy pp
collisions \cite{McLerran:2010ex,Praszalowicz:2011tc}. An onset of GS
in heavy ion collisions at RHIC energies has been reported in 
Refs.~\cite{Praszalowicz:2011rm}.

The proton seen at low Bjorken $x<x_{\mathrm{max}}$ is dominated
by a gluon cloud characterized by a typical scale $Q_{\text{s}}(x)$ which is
called saturation scale \cite{sat1,sat2,GolecBiernat:1998js}. 
The phenomenon of saturation (for a review see \cite{Mueller:2001fv,McLerran:2010ub})
appears due to the nonlinearities of parton evolution at small $x$ given
by so-called JIMWLK hierarchy equations \cite{jimwlk} which in the large
$N_{\rm c}$ limit reduce to the Balitsky-Kovchegov equation \cite{BK}.
These equations admit traveling wave solutions which explicitly exhibit
GS \cite{Munier:2003vc}. The appearance of GS in the prototype of the BK
equation obtained in double logarithmic approximation by Gribov, Levin and Ryskin
\cite{sat1} has been studied in Ref.~\cite{Bartels:1992ix}.
An effective theory describing the small $x$ regime 
is Color Glass Condensate \cite{sat1,sat2,MLV}.
For the present study the details of saturation are not of primary
interest, it is the very existence of the saturation scale which plays
the crucial role.

Recently it has been shown \cite{Praszalowicz:2012zh}
that  GS in DIS works very well up to relatively large $x_{\text{max}}\sim0.1$
(see also \cite{Caola:2010cy}). This should not come as a total
surprise since it is known that GS scaling extends well above the saturation scale
both in the DGLAP \cite{Kwiecinski:2002ep} and BFKL \cite{Iancu:2002tr} 
evolution schemes not only for the boundary
conditions that exhibit GS \cite{Caola:2008xr} to start with.
For $x<x_{\mathrm{max}}$ in
the first approximation the virtual photon cross-section $\sigma_{\gamma
^{\ast}p}$ in DIS is a function of a scaling variable $\tau = -q^{2}/Q_{\text{s}}%
^{2}(x)$ only, where $q^{2}$ is photon virtuality \cite{Stasto:2000er,GolecBiernat:1998js}. 
Similarly in hadronic collisions charged particle multiplicity in central rapidity exhibits GS
\cite{McLerran:2010ex}
\begin{equation}
\left.  \frac{dN}{dy d^{2}p_{\text{T}}}\right\vert _{y\simeq0}=\frac
{1}{Q_{0}^{2}}F(\tau)\label{GSinpp}%
\end{equation}
where $F$ is a universal dimensionless function of the scaling variable
\begin{equation}
\tau=p_{\text{T}}^{2}/Q_{\text{s}}^{2}(x).\label{taudef}%
\end{equation}
In this paper we shall address a question of applicability of
Eq.~(\ref{GSinpp}) beyond the central rapidity region. This is an important and
interesting question for the following reasons. First, Bjorken $x$'s of
scattering gluons that produce a particle of a given transverse momentum
$p_{\mathrm{T}}$ and rapidity $y$ read
\begin{equation}
x_{1,2}=e^{\pm y}\,p_{\text{T}}/W\label{x12}%
\end{equation}
and can be quite different for large rapidities (we shall assume positive $y$,
hence $x_{2}<x_{1}$). Here $W=\sqrt{s}$ denotes pp scattering energy. If
$x_{2}$ and $x_{1}$ are different then two different saturation scales emerge:
$Q_{\text{s}1}$ and $Q_{\text{s}2}$ \cite{Kharzeev:2001gp}. 
Whether this leads to violation of GS
depends on the form of the saturation scale (for discussion of different
forms of scaling variable see {\em e.g.} \cite{Beuf:2008mf} and references
therein). Second, and -- as we shall
argue below -- more importantly, when the larger Bjorken $x=x_{1}$ for some
$y$  leaves the domain of GS, \emph{i.e.} $x_{1}>x_{\text{max}}$,
violation of GS should appear due to the fact that one of the scattering
gluonic clouds becomes  dilute and quark degrees of freedom become important. We are
going to show in a model-independent way that this is what indeed does happen.

The theoretical formula for the multiplicity density in pp collisions is known
since 1981 \cite{Gribov:1981kg}:%
\begin{equation}
\frac{dN}{dy d^{2}p_{\text{T}}}=\frac{C}{p_{\text{T}}^{2}}%
{\displaystyle\int}
d^{2}\vec{k}_{\text{T}}\,\alpha_{\text{s}}\, \varphi_{1}(x_{1},\vec{k}_{\text{T}%
}^{2})\varphi_{2}(x_{2},(\vec{k}-\vec{p}\, )_{\text{T}}^{2}). \label{Nchdef}%
\end{equation}
Here $1/p_{\text{T}}^{2}$ corresponds to the $gg\rightarrow g$ cross section,
$\varphi_{1,2}$ are unintegrated gluon densities and $\alpha_{\text{s}}$
stands for strong coupling constant. Although in principle (\ref{Nchdef})
describes gluon production, one uses parton-hadron duality \cite{Dokshitzer:1991eq}
to argue that it can be also applied to particle multiplicities where constant
$C$ takes care of the hardonization effects.

At high energies and small $x$'s unintegrated gluon densities 
depend on $k_{\text{T}}$ through the ratio $k_{\text{T}}%
^{2}/Q_{\text{s}}^{2}(x)$ \cite{GolecBiernat:1998js}
-- which is an underlying mechanism behind
geometrical scaling.
Throughout this paper we shall use the following form of the saturation scale
\cite{GolecBiernat:1998js}:
\begin{equation}
Q_{\text{s}}^{2}(x_{1,2})=Q_{0}^{2}(e^{\pm y}p_{\text{T}}/(x_{0}\sqrt
{s}))^{-\lambda} \label{Qsat}%
\end{equation}
although other possibilities have been also discussed in the literature \cite{Beuf:2008mf}.
Parameters $Q_{0}\sim1$
GeV$/c$ and $x_{0}\sim10^{-3}$ depend on a particular model of gluon densities,
whereas exponent $\lambda$ can be extracted from the data in a
model-independent way. Indeed, by analyzing multiplicity spectra at CMS we
have found that for pp collisions $\lambda=0.27$ \cite{McLerran:2010ex}. 
Note that for $\lambda=0$
saturation scale is equal to $Q_{0}$ and scaling variable $\tau=p_{\text{T}%
}^{2}/Q_{0}^{2}$ is essentially transverse momentum squared in units of
$Q_{0}^{2}$.

There are many sources of possible violation of GS in multiplicity
spectra given by Eq.(\ref{GSinpp}).
First -- as already discussed -- one of the Bjorken $x$'s (\ref{x12}) can be
outside the GS domain. Second, gluons of very low $k_{\text{T}}$
will feel nonperturbative effects of the order of $\Lambda_{\text{QCD}}$.
Next, running of the strong coupling $\alpha_{\text{s}}$ will evidently
produce logarithmic corrections to GS. Finally, gluon fragmentation into
physical particles will introduce dependence on their masses, which should
induce GS violation in spectra of identified particles.

In this paper we shall concentrate on the first two effects which can be
studied by looking at rapidity dependence of the particle multiplicities in pp
collisions. For particles produced with $y\neq0$ Bjorken $x$'s (\ref{x12}) are
different and we have in fact two different saturation scales \cite{Kharzeev:2001gp}. 
Hence instead
of $F(\tau)$ in Eq.(\ref{GSinpp}) we will have some other function $\tilde
{G}(\tau_{1},\tau_{2})$. This by itself does not 
mean that GS is necessarily violated
since we can always rewrite $\tilde{G}(\tau_{1},\tau_{2})=G(\tau_{1},\rho)$
with $\rho=Q_{\text{s}1}/Q_{\text{s}2}$. For fixed rapidity $y$ ratio $\rho$
is constant for $Q_{\text{s}}$ of the form given by Eq.(\ref{Qsat}) and
universal dependence of multiplicity distribution upon scaling variable
$\tau_{1}$ is retained. Note that $F(\tau)=G(\tau,1)$.

\begin{figure}[h!]
\centering
\includegraphics[width=8.5cm,angle=0]{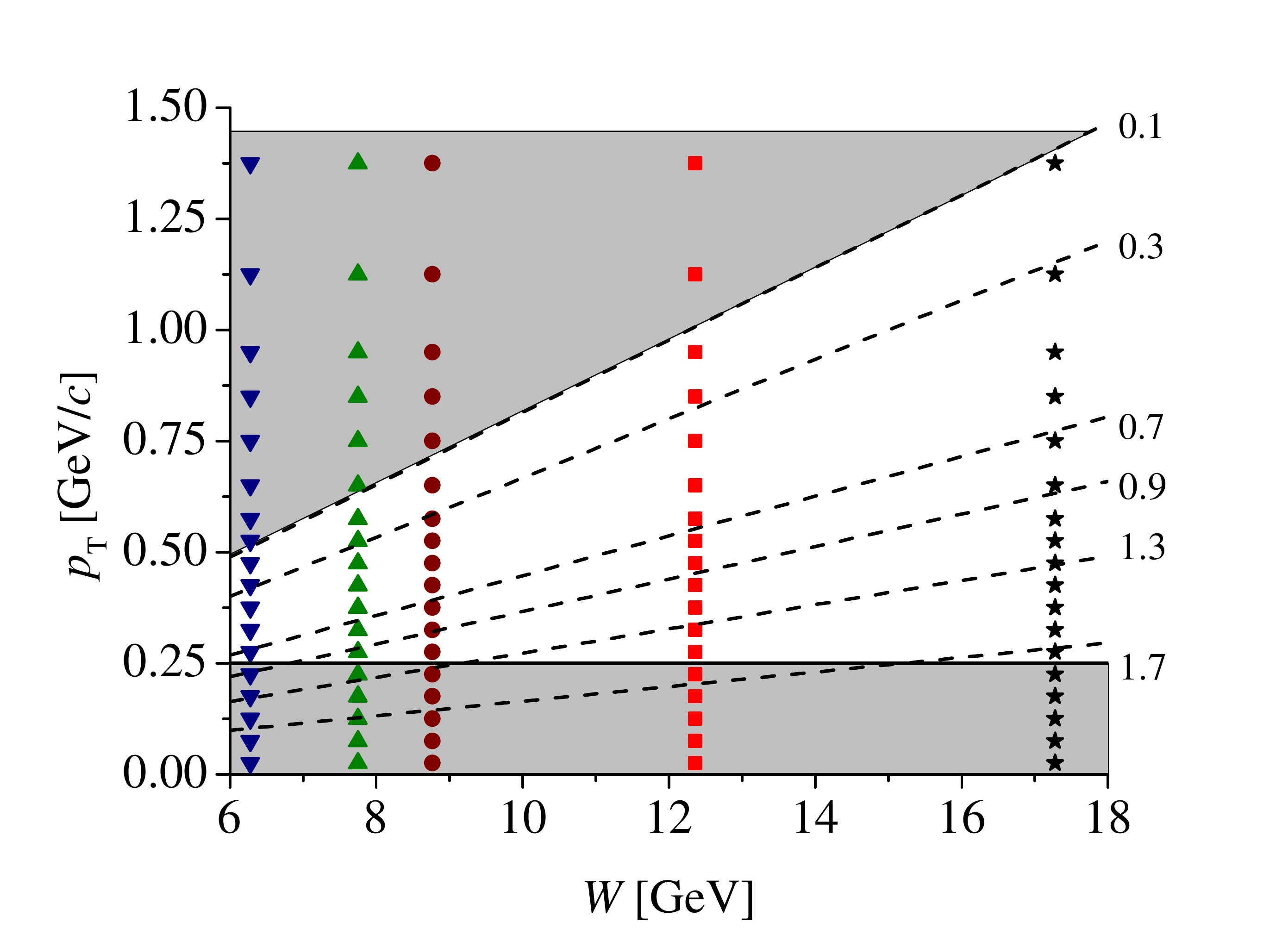} \caption{Kinematic
range of NA61/SHINE experiment in the $(W,p_{\mathrm{T}})$ plane. Data points are shown
as stars, squares, circles and triangles. Shaded areas denote regions where GS
is expected to be violated for $y=0.1$. For other rapidities (marked at the
r.h.s. of the plot) GS violation takes place above the respective dashed
line.}%
\label{kinrange}%
\end{figure}

On the other hand by looking at the spectra with increasing $y$ one can
eventually reach $x_{1}>x_{\mathrm{max}}$ and GS violation should be seen.
Luckily we have at our disposal very recent
pp data from the NA61/SHINE experiment at CERN \cite{NA61}
where particle spectra at different rapidities $y=0.1\div3.5$ and five different
energies have been measured for $p_{\text{T}}=0.025\div1.375$ GeV$/c$. For the
present analysis we shall use negative pion spectra, which has an advantage
that one avoids possible GS violation due to particle masses. The only problem
with this piece of data is that the scattering energies are rather low:
$W_{1_,\ldots_,5}=17.28,\;12.36,\;8.77,\;7.75$, and $6.28$ GeV. Nevertheless, as we
will show below, Bjorken $x$'s involved are small enough for GS to be present.
This is mainly due to the before-mentioned fact that in DIS $x_{\text{max}%
}\sim0.1$ \cite{Praszalowicz:2012zh,Caola:2010cy}; 
for the purpose of the present analysis we shall take
$x_{\text{max}}=0.09$. Then GS should be seen if $x_{1}<x_{\text{max}}$,
\emph{i.e.} for%
\begin{equation}
p_{\text T}< p_{\text{Tmax}}(W,y)=x_{\text{max}}W\,e^{-y}. \label{ptmax}%
\end{equation}
On the other hand transverse momenta of produced gluons should be larger than
some nonperturbative scale $\Lambda$:%
\begin{equation}
p_{\text T}>p_{\text{Tmin}}=\Lambda\label{ptmin}%
\end{equation}
for which we take $\Lambda=250$ MeV \cite{footnote}. 
Having fixed the kinematical range where GS
should be present, we can have a look at kinematical range of NA61/SHINE spectra
shown in Fig.~\ref{kinrange}. 
The shaded area below $p_{\text{Tmin}}=\Lambda$ is excluded for
all rapidities. The shaded region in an upper part of the plot is excluded for
$y=0.1$, and dashed lines represent $p_{\text{Tmax}}(W,y)$ as functions of $W$
for different rapidities $y$. Therefore for given rapidity only points below
these lines should exhibit GS. Experimental points for different energies are
represented by stars (17.28 GeV), squares (12.36 GeV), circles (8.77 GeV), and
triangles (7.75 and 6.28 GeV). One can see from Fig.~\ref{kinrange} 
that for $y=0.1$ 
 GS region extends towards the smallest NA61/SHINE energy. 
 This is due to the fact that $x_{\rm max}$ is as large as 0.09.
 By increasing
$y$ we see that some points fall outside the  GS window and finally for
$y\gtrsim1.7$ no GS should be present in NA61/SHINE data. 
One may note that both
$x_{\text{max}}$ and $\Lambda$ together with exponent $\lambda$ could be
determined in a self-consistent way from the NA61/SHINE data. This is beyond the
scope of the present note and will be presented elsewhere. Therefore our
analysis has to be considered mostly as a qualitative one.

\begin{figure}[h!]
\centering
\includegraphics[width=8.5cm,angle=0]{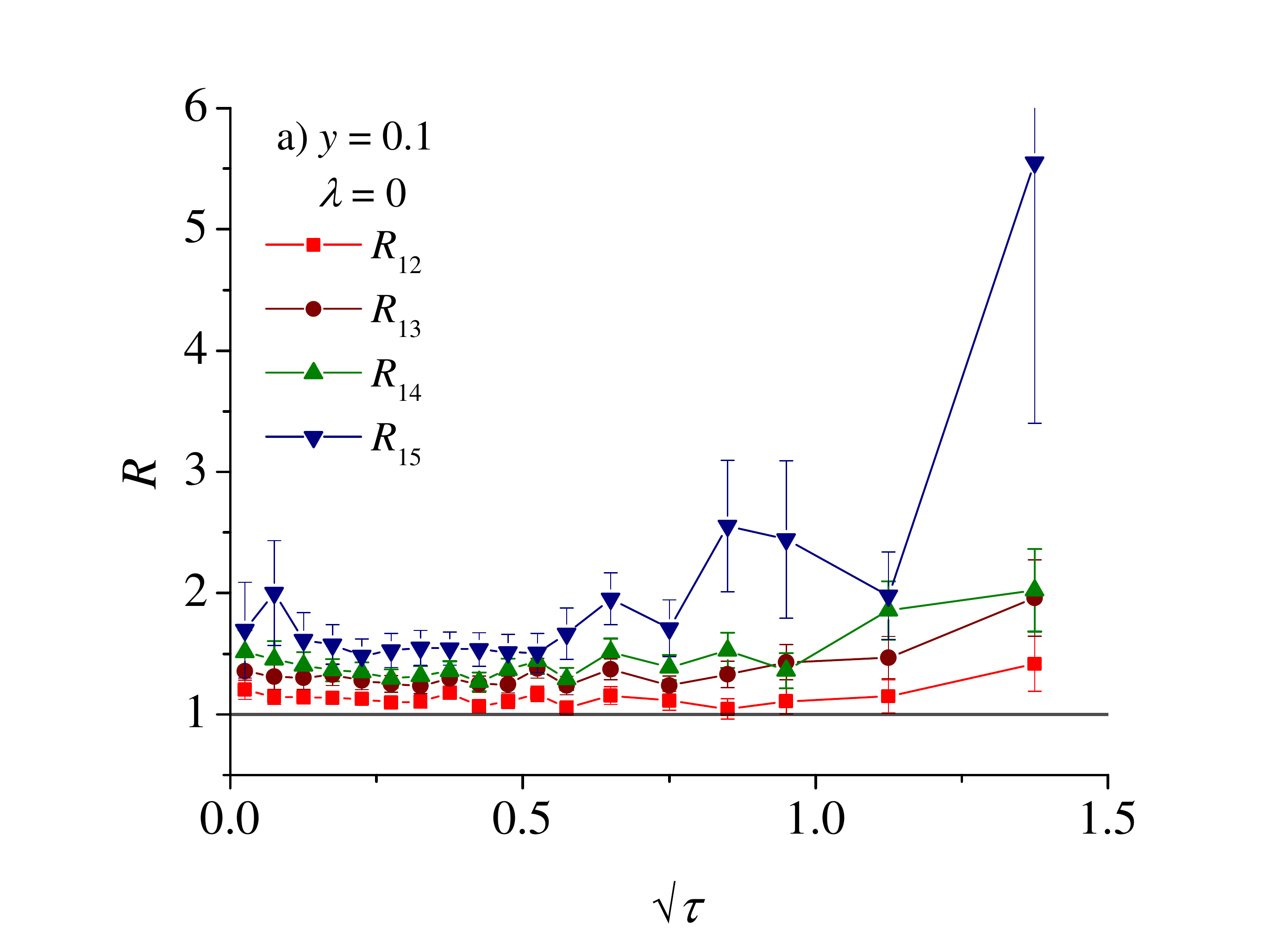}
\includegraphics[width=8.5cm,angle=0]{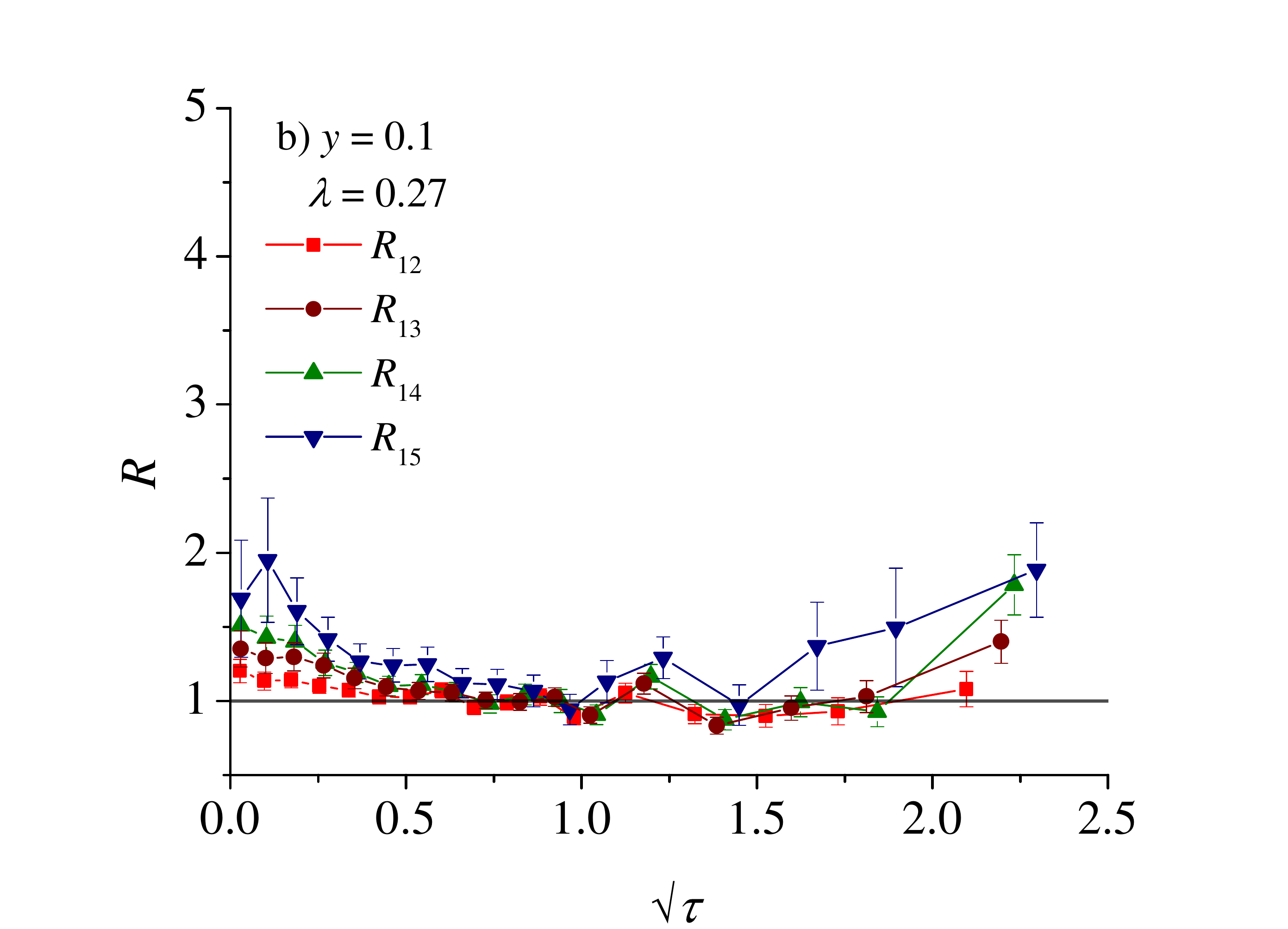} \caption{Ratios $R_{1k}$
as functions of $\sqrt{\tau}$ for the lowest rapidity $y=0.1$: a) for
$\lambda=0$ when $\sqrt{\tau}=p_{\mathrm{T}}$ and b) for $\lambda=0.27$ which
corresponds to GS.}%
\label{y01}%
\end{figure}

Limits (\ref{ptmax}) and (\ref{ptmin}) define  GS window $\tau_{\text{min}%
}<\tau<\tau_{\text{max}}$ in scaling variable $\tau$%
\begin{align}
\tau_{\text{max}}  &  =e^{-2y}x_{\text{max}}^{2}W^{2}/Q_{0}^{2}%
\,(x_{\text{max}}/x_{0})^{\lambda},\nonumber\\
\tau_{\text{min}}  &  =e^{\lambda y}\Lambda^{2}/Q_{0}^{2}\,(\Lambda
/(x_{0}W))^{\lambda}. \label{taumaxmin}%
\end{align}
We see from (\ref{taumaxmin}) that for fixed $W$, $\tau_{\text{max}}$ falls
exponentially with rapidity $y$ much faster than $\tau_{\text{min}}$ is rising
with $y$. When $\tau_{\text{min}}\simeq\tau_{\text{max}}$ window for GS
closes, and we get from (\ref{taumaxmin}) that the larger $W$ the larger $y$
when this happens. For fixed $y$ the GS window is growing with $W$.

Now we come to the formulation of a model-independent criterion that allows one to
asses whether GS is present or not. To this end we follow 
Refs.~\cite{McLerran:2010ex,Praszalowicz:2012zh} 
and define
ratios of multiplicities%
\begin{equation}
R_{ik}(\tau; y)=\frac{dN}{dy d^{2}p_{\text{T}}}(W_{i},\tau; y)/\frac{dN}{dy
d^{2}p_{\text{T}}}(W_{k},\tau; y)\label{Rik}%
\end{equation}
which, according to (\ref{GSinpp}), should be equal to unity if GS is present.
Since GS window is the largest for $W_{1}=17.28$ GeV,  we form four
ratios $R_{1k}$ with $k=2,\ldots,5$ where index $k$ refers to NA61/SHINE energies
mentioned above. In Fig.~\ref{y01} we plot ratios $R_{1k}$ as functions of
$\sqrt{\tau}$ for the lowest rapidity $y=0.1$ first for $\lambda=0$,
\emph{i.e.} essentially as functions of $p_{\text{T}}$, and second for
$\lambda=0.27$. We see that for $\lambda=0$ all ratios $R_{1k}$ are larger
than $1$ and rise with $\tau$ up to $R_{1k}\sim1.5\div6$. In contrast, for
$\lambda=0.27$ for all ratios $R_{1k}$ there exists a domain in $\sqrt{\tau}$
where, within the errors, they are equal to 1. We see that $\tau_{\text{min}}$
for $W_{2}=12.36$ GeV is approximately 0.6 while for $W_{5}=6.28$ GeV it is
0.8 or so. On the other hand for $W_{5}$ the GS windows closes at $\sqrt{\tau
}\simeq1.5,$ whereas for $W_{2}$ it extends further than 2.1. This is in
qualitative agreement with limits (\ref{taumaxmin}), and as we will see in the
following, this trend continues once we increase $y$.

\begin{figure*}[h!]
\centering
\includegraphics[width=8.5cm,angle=0]{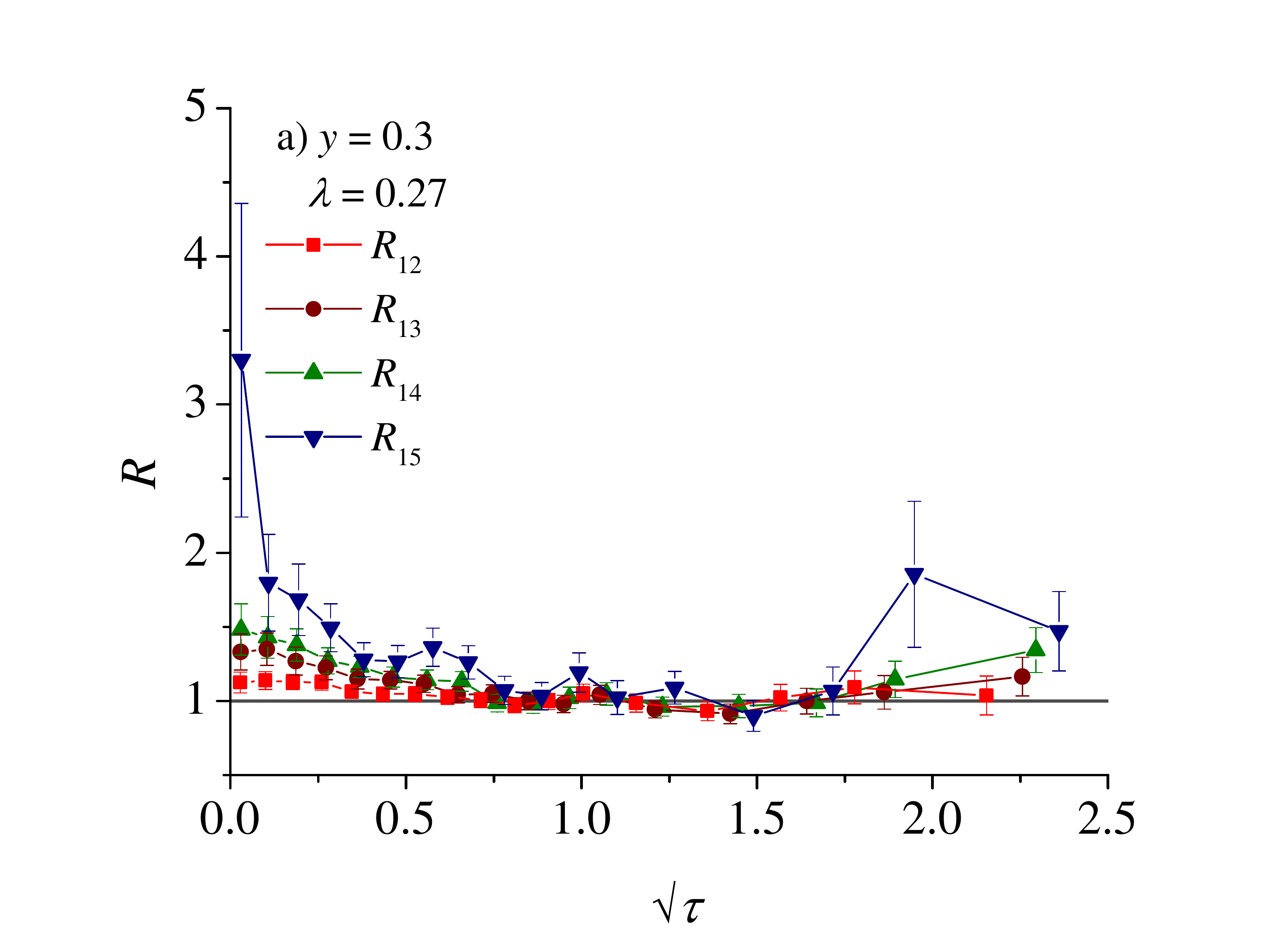}
\includegraphics[width=8.5cm,angle=0]{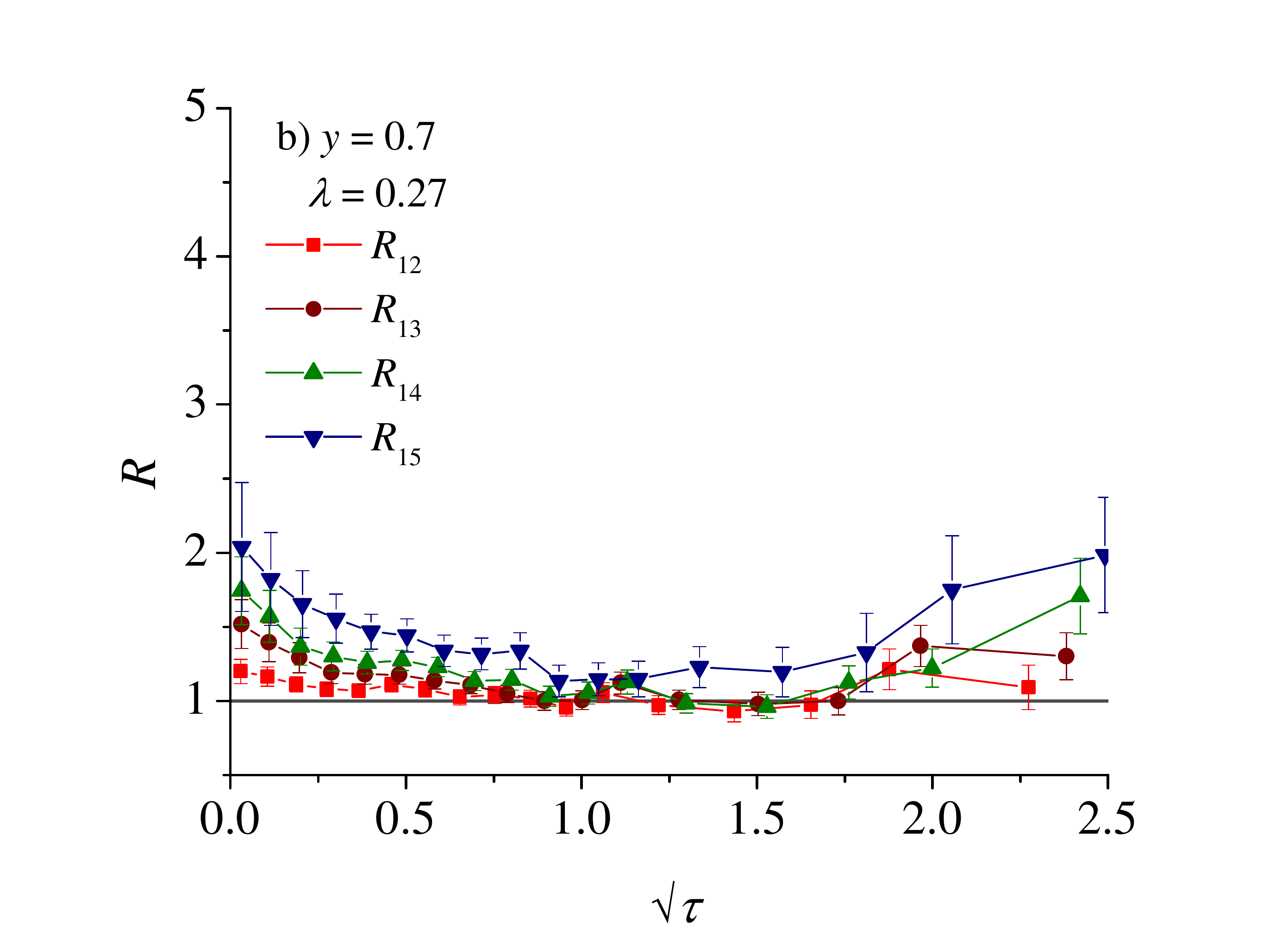}\\%
\includegraphics[width=8.5cm,angle=0]{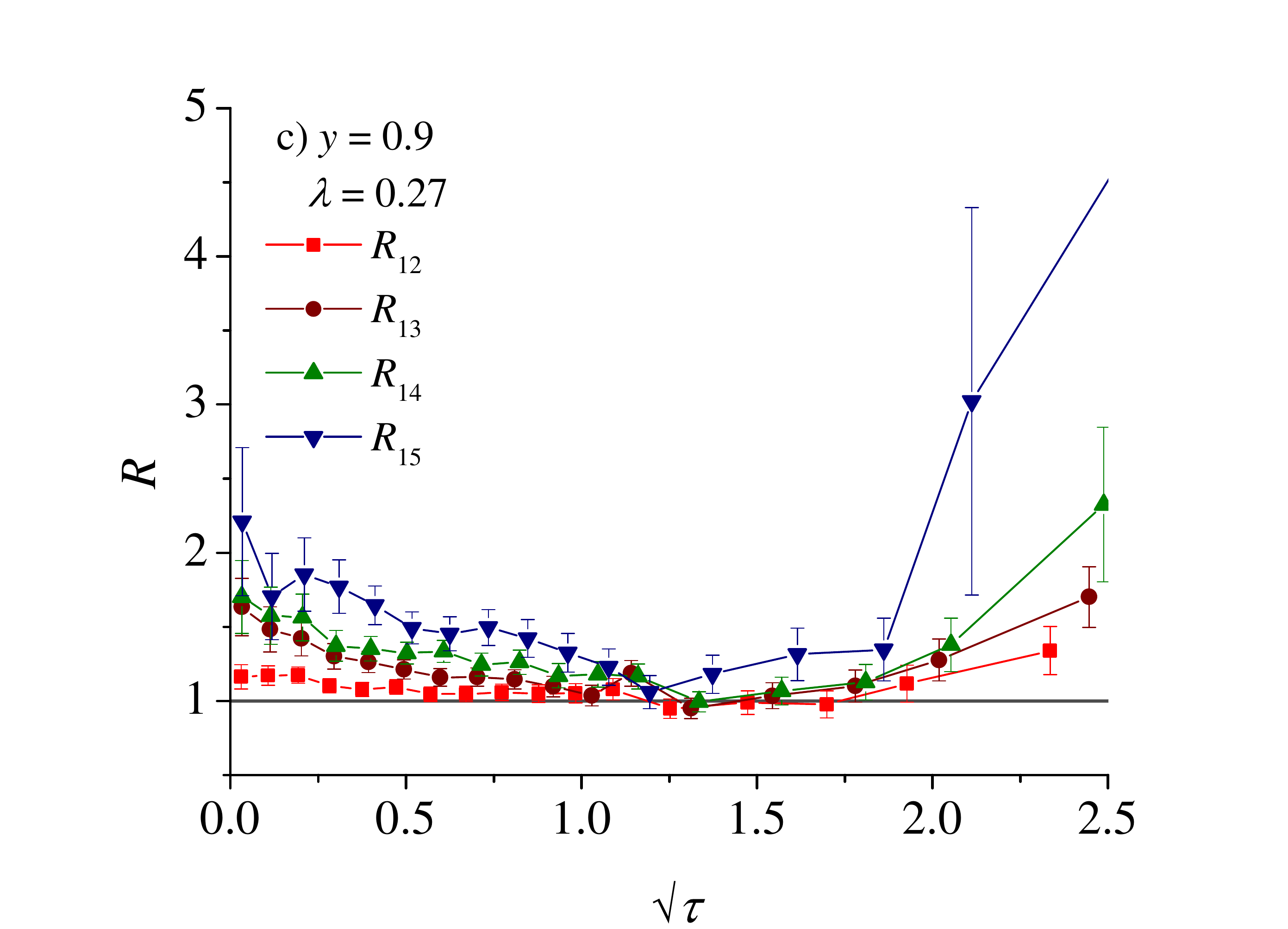}
\includegraphics[width=8.5cm,angle=0]{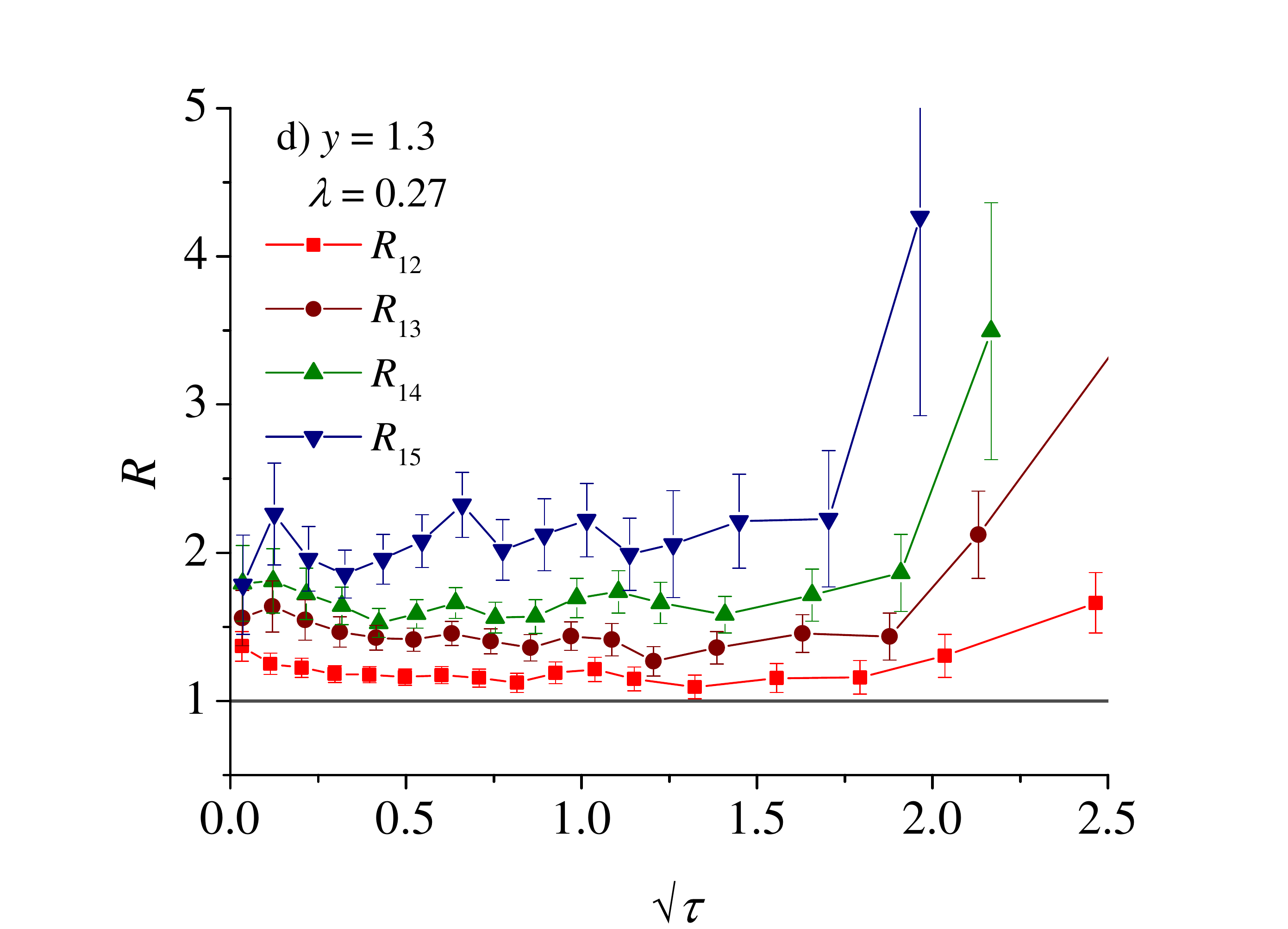} 
\caption{Ratios $R_{1k}$
as functions of $\sqrt{\tau}$ for $\lambda=0.27$ and for different rapidities
a) $y=0.3$, b) $y=0.7$, c) $y=0.9$ and d) $y=1.3$. With increase of rapidity,
gradual closure of the GS window can be seen.}%
\label{ys}%
\end{figure*}

Looking at Fig.~\ref{kinrange} we see that there is no qualitative difference
between rapidities $0.1$ and $0.3$, and this is what indeed can be seen in
Fig.~\ref{ys}.a. However, already for $y=0.7$ the lowest energy data are no
longer in the scaling region, and we see this clearly in Fig.~\ref{ys}.a.
Further on, for rapidity $y=0.9$ the second lowest energy data leave the GS
window, and this is again nicely substantiated by the behavior of $R_{14}$
shown in Fig.~\ref{ys}.c. Finally for $y=1.3$ none of the tested energies
$W_{2}\div W_{5}$ should exhibit GS, and this is clearly confirmed by
Fig.~\ref{ys}.d. Note that in all these cases the reference points at
scattering energy $W_{1}$ remain in the GS window,  therefore scaling
violations of ratios $R_{1k}$ should be attributed to the points at lower
energies $W_{k}<W_{1}$.

It is somewhat surprising that relatively low energy NA61/SHINE data do show GS at
all. This is due to the fact that GS scaling works up to much higher $x$'s
than originally anticipated: $x_{\text{max}}\sim0.1$. This has been confirmed
by direct analysis of DIS in Ref.~\cite{Praszalowicz:2012zh} 
and also indirectly by the present
study. 

In this paper we have explored NA61/SHINE \cite{NA61} 
negative pion spectra in pp collisions taken at
different rapidities $y$. 
 In order to study GS in multiplicity distributions we have formed
ratios (\ref{Rik}) that should be equal to 1 for scaling variable $\tau$
within the GS window (\ref{taumaxmin}).
We have argued that behavior of ratios $R_{1k}$ with increasing
$y$ is in qualitative agreement
with the following picture of particle production in hadronic collisions. For
low $p_{\text{T}}$'s  GS is violated due to the existence of the nonperturbative
energy scale $\Lambda$. For high transverse momenta, the larger of two Bjorken
$x$'s, namely $x_{1}$, crosses the maximal $x$ for which GS is present,
$x_{1}>x_{\text{max}}$, and GS is again violated. For rapidities larger than
1.3, the window of GS closes for all energies $W_{k}$ but $W_{1}$, and no GS is
seen in ratios $R_{1k}$. This behavior of GS for different energies with
varying $y$ is strikingly in qualitative agreement with the kinematical
constraints shown in Fig.\ref{kinrange} and more quantitatively with the range
of  the GS window given by Eq.(\ref{taumaxmin}).

\newpage

Our aim here was to point out that model-independent analysis of the
multiplicity distributions at different rapidities can provide an interesting
insight into the production mechanism of low and medium $p_{\text{T}}$ pions
from the point of view of gluon saturation and geometrical scaling. One
obvious reservation is that for higher rapidities one can force ratios
$R_{1k}$ to approach unity at the expense of adjusting $\lambda$. However,
very soon $\lambda$ gets bigger than $1$ and for rapidities of the order of 2
it has to be bigger than $3.$ The quality of such "geometrical scaling" is very
poor, and -- more importantly -- there is no physical picture supporting such
big values of $\lambda$. This is one of the reasons for which one should
perform a more quantitative analysis of GS in pp collisions, which will be
presented elsewhere.

Let us finish by a remark that experimental data at high 
LHC energies  would be best suited for an analysis outlined
in this paper. For the LHC energies GS closes for very high 
rapidities up to $y \sim 8$.
Since one needs data
at different rapidities, perhaps the best
detector for this purpose would be LHC-b.  

\bigskip
The author wants to thank Marek Ga{z}dzicki and Szymon Pulawski for
the access to the NA61/SHINE data, and Andrzej Bia{l}as  and Marek Ga{z}dzicki
for careful reading of the manuscript and remarks. This work was supported by 
the Polish NCN  Grant 2011/01/B/ST2/00492.

\end{document}